# Making sense of studying physics

Åke Ingerman[1] & Shirley Booth[2]
[1]Department of Physics, University of the Western Cape, Cape Town, South Africa
[2]Department of Education, Lund University, Sweden
aingerman@home.se; Shirley.Booth@pedagog.lu.se

With an investigation into how students in a physics Master of science program make sense of their whole first year study experience in one of the years following a programme reform, we try to offer insights and advice to consider when involved in similar changes. We conclude and argue that by trying to support the students' in making sense of their studying in terms of physics and empowering them in their understanding of the nature of physics and their study situation, both the potential for productive, meaningful and positive learning of physics as something which is relevant in the students life, and the way in which the students analyse and make sense of things they encounter, both physically and academically, are advanced.

**Introduction and background**

In South Africa, there is a strong movement to curriculum change and the implementation of new ideas, in physics as well as in most other subjects, to adapt to the changing situation there. These changes and the design of new programmes and courses are generally made with teaching as the sole starting point. However, even if a new design might look complete from the teachers' perspective, the ways in which the students make sense of the content and their study situation in this new programme may be radically and essentially different from those intended. With an investigation into how students made sense of their first year of studying physics at a Swedish top engineering university (Chalmers University of Technology) a few years after the implementation of a programme reform, we will try to offer some insights and advice to consider when involved in similar changes. The programme which the students were taking is a four-and-a-half year Master of science-programme in physics. It is structured around a large number of content specific courses, primarily in physics and mathematics, but also having a substantial part of general engineering subjects. The students taking the programme have generally been very successful and hard working students during high school, but their results in the programme range from success to complete failure.

**Phenomenographic approach**

The study (which is fully reported in Booth & Ingerman, 2002) was carried out with a predominantly phenomenographic approach (Marton, 1981; Marton & Booth, 1997). This implies that we were interested in variation in the ways in which the students experienced their first year of studying physics with respect to its content and structure. Data was collected primarily through interviews with 20 students at some time in their second year of studies, selected to represent a cross-section of success.

In line with the phenomenographic approach, the interviews (and the transcriptions of them) are seen as forming a 'pool of meaning' in which the variation in ways of experiencing the phenomenon of interest is to be seen. By reading the interviews repeatedly, now as expressions of individual students, now as series of extracts related to specific issues, we delved more and more deeply into the meaning of 'studying physics' as seen by the students. Categories were formed and reformed; extracts from interviews were sought to support and give substance to the categories; and logical and empirical links between categories were explored. The aim was to offer a hierarchy of empirically grounded and logically consistent categories of description, which capture the essence of the whole experience and reveal the essential variation in the structure of that experience. This result of this process is described in the next section.

**Ways of making sense of studying physics in the first year of study**

The outcome of our analysis is six distinct, qualitatively different categories, describing different ways of making sense of studying physics, ranging from focusing on the demands and elements of the study situation as such, trying to cope with unrelated fragments of physics knowledge, to making sense of courses starting from the physics itself, constructing a physics knowledge object in the context of 'real world' physics.

We introduce the term "knowledge object" (Entwistle & Marton, 1994) in two different forms: the "physics knowledge object" refers to the developing body of knowledge of physics and the structure and complexity it takes on as it emerges from the courses in mathematics, physics and engineering, while the "study knowledge object" refers rather to the structure and complexity of the approach to studying courses. They can be seen as distinctly different senses being made of the study situation.

**Table 1**: Summary of the different ways of making sense of studying physics in the first year of study.

| Category | Description |
|---|---|
| Courses are identified with the study situation | Here the engineering physics programme has been experienced as a discrete set of courses, a means to the end of a degree. These are related to authority, i.e. teachers and tradition, and common features, such as the ways in which courses were organised. |
| One course is seen as a prerequisite for another course | Courses are now related to their content to the extent that a preordained, correct sequence of acquisition of knowledge fragments is assumed. A 'red thread' is sought in terms of needs and demands. Authority for the thread — content and structure — is still the domain of teachers and tradition. |
| One course is seen as being useful in other courses | Courses now support one another, but they still are necessarily arranged in a specific order. Reference is made to the knowledge fragments that constitute the courses, which mesh into one another, course-to-course. |
| Courses are related through mutual illumination | Here is to be found sense-making for the first time. Courses now lend meaning to each other and understanding in an earlier course can be found in a later course. There are now networks that mesh and unmesh, knowledge fragments might be grouped together in different ways and offer different perspectives. There is a dynamic in what is focal or non-focal, and thematic or non-thematic. The Physics that is constituted takes on a dynamic form and begins to resemble a 'physics knowledge object' rather than a 'study knowledge object'. |
| Courses fit together into an adaptable whole | The courses are seen as constituting parts of a whole, and the strict ordering structure of the educational programme knowledge content is broken apart. An internal dynamic enables a picture to develop which is different on different occasions, depending on what aspects are brought into focus. |
| Courses in physics come into physics of the world | The borders between courses are erased, a physics knowledge object is constituted, physics and the physics world are one with the knower. |

**Discussion**

Clearly, the programme is very important for framing the students' experiences. It can justifiably be said that for the students, the physics programme was the major factor in creating the conditions for learning they made sense of in different ways. Even though we cannot say much about individual courses and teachers, it is clear that the programme as a whole, its organisation and reality had an overwhelming impact. From this observation, we argue that any programme that is organised as this one is, as a set of courses given by subject specialists, (and degree programmes mostly are), should have as an overriding goal that the students come to see the subject matter as a related whole, and that this provides them with ways of seeing and coping with a world as yet unknown. Or to state the goal differently, that the students should make sense of their learning primarily in terms of physics. We can describe as the goal that the students should, as a result of their extensive physics studying, create a physics knowledge object – a tightly integrated, independent and flexible body of physics knowledge – which can function as a lens through which one can experience and analyse both academic physics problems and everyday real world physics contexts.

Some of the students we interviewed could indeed be characterised as having formed such a physics knowledge object. However, that was far from true for the majority of students. We propose that two major obstacles in formulating and implementing such goals is firstly the fact that the goals are unarticulated and allowed to remain abstract (by the collective of teachers), and secondly that the teachers' perspective dominates while the students' perspective is taken for granted or ignored. To articulate goals of this kind and to explore how they can be manifested in practice in teaching and creating learning environments, can only be achieved by a collective engagement and effort among teachers and students to discuss them, negotiate around them and try them out in cooperation.

To articulate goals might seem quite straight-forward (even though it often turns out to be more complicated than you expect), but what does it mean to take the students' perspective of studying and learning physics seriously into account? We mean that it is a stand which has implications on many different levels. To take students' perspective into account is not restricted to, or even mainly, a question of listening to their opinions. Firstly, it is a question about continuously exploring and being aware of different ways of seeing physics phenomena and contexts, that is to strive to see the object of learning from the students' perspective. Secondly, it is a question which also has roots in the view of the nature of physics knowledge, that is to be able to (and to actually) see (and supporting students to see) further than the physics at hand, to be explicitly aware of the very nature of how physics knowledge is constructed and how we want it to be understood. Thirdly, it is also a question of communicating, appreciating and empowering: communicating with the students on their understanding and their study situation; appreciating the students' ways of understanding the physics; empowering the students to analyse their own physics understanding and their study situation; and empowering them to reflect on, articulate and discuss both their understanding and features of their situation, and thus potentially find new ways of understanding and address problematic features of their study situation.

Addressing the implications of our results and reasoning are, we must regretfully admit, not instantaneously done, but takes some consideration, cooperation and effort. The path to progress is not to be walked by either students or teachers alone. But it is we, as teachers, who are morally responsible for opening up the path. For the students, it is important that they are supported in taking control of their studies and becoming aware of why and how they make sense of and manage their study situation. At Chalmers University of Technology, we have during recent years been offering a course focusing on these issues based on our results, called 'Towards better leaning'. In the next section, we will briefly outline the course as well as the considerations made and the experiences gained. For the teachers, we see that the efforts of individual teachers would dramatically benefit if supported by a collegiality of teachers, openly discussing how, both on the general programme level and in the local course context, to bring – to the highest possible degree – the physics, its structure and its internal relations within and between courses, into the students' awareness. In the section 'Conceptual physics' we will outline some experiences drawn from the conceptual physics courses, which are being given at Uppsala University in Sweden, and, in the South African context, at the University of the Western Cape. In these courses the teachers have developed an 'internal' form of collegiality, as well as ways of addressing issues of being aware of the students' different ways of experiencing physics phenomena and the nature of physics, and empowering the students in their studying of physics. In the final section, we will round off this article by discussing possible ways of developing a collegiality of teachers.

**Towards better learning**

The starting point for the course was the conjecture that students through articulation and reflection around their learning can become more aware about their own learning, and make more conscious choices about their study situation. The course was organised as a set of meetings during the first semester, between the students in groups of 6 to 8 and a supervisor, revolving around a series of assignments and discussions based on and relating to those assignments. Following this string of assignments, the students gradually were led into systematically exploring their own study situation, considering wider and wider perspectives on it, and trying out ways of becoming/being the most important actor in that situation. Starting out with self-observation, they proceeded to explore other students' perspectives and a teacher's perspective, and then they returned to set goals and objectives for their own situation.

The assignments took their individual starting points in an observation the students should make concerning their own study situations, and often had both abstract and more practical objectives. The core was often about relating action in practice to underlying perspectives, motives and alternative actions. In practice this meant that the students were required to report in the form of a (short) reflective essay, in which they were expected to use 'academic argumentation' –that is, start with some observation, engage in some kind of rational reasoning and come to some conclusion or suggestion for action or change. On the basis of their assignments, the group discussed the topic (as well as others related to it) and on the basis of their essays, the students got feedback from the supervisor, similar to what in the academic world is called peer review. In other words, in the feedback we tried to recognise the students' views and opinions, but we urged them to (and supported them in) problematizing the things they might be taking for granted.

We will not discuss all of the assignments here, but just briefly outline two of them, the 'study diary' and the 'interview with a teacher'. The students meet the study diary assignment twice, once in each term. Over a period of a week, they have to write down all study activities they are involved in, including lectures. The first time, the aim is to make the students get a realistic view of what and how they are studying, and whether they are 'effective' when doing it. The second time, they first have to articulate their (qualitative and quantitative) goals, and the assignment is to evaluate their activities against their goals. In the interview with a teacher, their assignment is to meet and discuss with one of the teachers they have met so far in one of their courses, to get some insight into their view on teaching and learning, and critically evaluate it. More details around the course and a more elaborated discussion of its ideology can be found in Ingerman *et al.* (2003).

The course is in its essence about empowerment of the students, to enable them to analyse and be creative in being students, changing the way in which their study situation is delimited and set up by the teachers. In its prolongation, it also has implications and potential for changing the ways in which the teachers understand what teaching is about, when interacting and entering into a dialogue with the empowered students.

**Conceptual physics**

The Conceptual physics-courses at University of the Western Cape and at Uppsala University are slightly different from most courses in physics. Apart from the physics content, there is an additional focus on reflection on learning, and supporting the students in becoming aware of their own learning. In comparison with the 'Towards better learning' course, this is integrated much more tightly with the course content, the physics. The reflection is encouraged through in-class discussion and reflective assignments.

Within the course, issues 'outside' normal physics, like beliefs around the nature of physics knowledge and scientific inquiry, are also addressed, for example through discussions and essays aimed at such issues. In particular, this is framed by the teachers to bring out contradicting or puzzling physics situations with philosophical undertones for discussion (in the whole class or in smaller group tutorials), or to ponder in an essay.

Another important strand is the ambition (and practice) of relating physics concepts and physics ways of organising reasoning and knowledge to everyday life and the world view *of the students*. That is, to try to bring out the everyday meaning of physics concepts on the one hand (e.g. mechanics when driving a car) and analysing everyday situations in physics terms on the other (e.g. scrutinizing the news and the claims made in newspapers). In general, this means talking and discussing physics in wider contexts, such as social (e.g. what implications has physics knowledge in the context of particular political questions), historical (e.g. what disputes have physicists been involved in over physics concepts which today look as if they were set in stone) and environmental contexts (e.g. issues of the spread of pollution and nuclear radiation risks). Differently framed, these activities fall under the heading of trying to achieve the goal of empowering students to think critically about phenomena related to physics, in the academic world and beyond.

To teach and develop the course described, a certain kind of discourse has evolved between the teachers involved, as an expression of an approach to their common teaching obligation, which encompasses a discussion of student learning (and goals of learning) in physics, and allows them to deal articulately with issues otherwise taken for granted. This discussion is the essence of what we would like to describe as a scholarly collegiality. How did that collegiality of scholarly approach to teaching (Boyer, 1990; Kreber, 2001) develop and how does it manifest itself in practice? Here, it has developed out of teaching in a team, which problematizes various matters of concern, and acts accordingly, taking into account research on students approach to physics learning and on students' conceptions of particular physics concepts. It manifests as systematically discussing the students' views on learning, identifying difficult concepts, what is difficult with those particular concepts, as expressed in class (lecture and tutorials) and in discussions with individual students. It includes inventing and trying out new ways of supporting students to change their views on learning and in this case also doing research about things which emerged as puzzling (see e.g. Linder & Marshall, 1997; 1998)

The course was conceptual, which means that it focused on the concepts rather than on the mathematical calculations. However, the features of the course outlined here is not tied to the conceptual nature in particular, but rather is an expression of a certain way of seeing teaching physics, which we regard as an example of taking the student's perspective seriously.

**Collegiality of teachers**

Unfortunately, the examples we have discussed in the previous two sections are isolated; they lack embeddedness in a 'college of teachers[1]'. Whole departments and/or the college of teachers teaching in a programme need to participate and engage in developing a discourse in which students' learning can be discussed and goals and approaches to achieving those goals can be discussed and developed. That is, to incorporate elements of students' learning in physics into the normal production of physics knowledge, and employing that knowledge in a scholarly approach to teaching. We see three important directions in which to go simultaneously to develop a scholarly collegiality on a departmental level:

We need to embark on a discussion aimed at exploring students' experiences in our departments. To our aid there are simple means as talking and listening to students one by one, using tools as the study diaries to get a more broad input on how students handle their study situation and the general student learning research literature.

---

[1] Where 'college' is being used in the sense of *"a society of persons joined together generally for literary or scientific purposes, and often possessing peculiar or exclusive privileges"* (Chalmers English Dictionary)

We need to develop in the college a shared and detailed awareness and understanding about different ways of understanding physics concepts. To aid us there is the educational research on students understanding of certain concepts, misconceptions and conceptual change. And most importantly, every teacher can explore, in a scholarly way, ways of understanding the main concepts they are teaching, of course with the aid of students.

Finally, we need to engage in a creative dialogue with students, empowered to analyze, articulate and discuss their study situation to deal with the problems we meet in the present, appreciate the good things that are done and decide on the path into the future.

**Acknowledgements**

We thank Delia Marshall and Cedric Linder, with whom we have had many enlightening and interesting discussions around the 'Conceptual physics' courses at University of the Western Cape and Uppsala University. Åke Ingerman has been supported financially by The Swedish Foundation for International Cooperation in Research and Higher Education (STINT).

**References**

Booth, S: & Ingerman, Å. (2002). Making sense of Physics in the first year of study. *Learning and Instruction*, 12, 493-507

Boyer, E.L. (1990). *Scholarship Reconsidered: Priorities of the Professoriate*. Princeton, N.J.: The Carnegie Foundation for the Advancement of Teaching

Entwistle, N. & Marton, F. (1994). Knowledge objects: understandings constituted through intense academic study. *British Journal of Educational Psychology*, 64, 161–178.

Ingerman, Å., Carling, K., & Booth, S. (2004). The paradox of studying effectively – Supporting first year physics engineering students in their learning situation, *manuscript*

Kreber, C. (Ed) (2001) *Scholarship revisited: perspectives on the scholarship of teaching*. San Fransisco: Jossey-Bass

Linder, C.J. & Marshall, D. (1997). Introducing and evaluating metacognitive strategies in large-class introductory physics teaching. In C. Rust. and G. Gibbs (Eds) *Improving student learning through course design*. Oxonian Rewley Press: Oxford.

Linder, C.J. & Marshall, D. (1998). Linking physics students' development as independent and reflective learners with changes in their conceptions of science. In C. Rust (Ed) *Improving students as learners*. Oxonian Rewley Press: Oxford.

Marton, F. (1981). Phenomenography—describing conceptions of the world around us. *Instructional Science*, 10, 177–200.

Marton, F. & Booth, S. (1997). *Learning and Awareness*. Mahwah NJ: Lawrence Erlbaum Ass.